\documentclass[aps,prb,10pt,twocolumn,superscriptaddress]{revtex4-1}

\usepackage{amsmath, amssymb}    					
\usepackage{graphicx}   							
\usepackage{xcolor}     							
\usepackage{hyperref}   							
\hypersetup{colorlinks=true,allcolors=blue}			
\usepackage{textcomp}								

\newcommand{\ddt}[0]{\frac{\partial}{\partial t}}
\renewcommand{\t}[1]{\textrm{#1}}
\newcommand{\nn}[0]{\nonumber\\}

\newcommand{\mbf}[1]{\mathbf{#1}}
\renewcommand{\k}[0]{\mathbf{k}}
\newcommand{\K}[0]{\mathbf{K}}

\renewcommand{\r}[0]{\mathbf{r}}
\newcommand{\R}[0]{\mathbf{R}}

\newcommand{\B}[0]{\mathbf{B}}

\newcommand{\up}[0]{\uparrow}
\newcommand{\down}[0]{\downarrow}
\newcommand{\Jsd}[0]{J_{sd}}
\newcommand{\Jpd}[0]{J_{pd}}
\newcommand{\NMn}[0]{N_\t{Mn}}

\newcommand{\omMn}{\omega_{\t{Mn}}}
\newcommand{\ome}{\omega_{\t{e}}}

\newcommand{\gel}{g_\t{e}}

\newcommand{\gMn}{g_\t{Mn}}

\newcommand{\bs}[1]{\boldsymbol{#1}}

\newcommand{\me}{m_{\t{e}}}

\newcommand{\etae}{\eta_{\t{e}}}
\newcommand{\etah}{\eta_{\t{h}}}

\newcommand{\ud}{{\uparrow/\downarrow}}
\newcommand{\du}{{\downarrow/\uparrow}}

\newcommand{\omsf}[0]{\omega_\mathrm{sf}}
\newcommand{\ph}[1]{\phantom{#1}}

\begin{document}

\title{Spin dynamics of hot excitons in diluted magnetic semiconductors with spin-orbit interaction}
\author{F. Ungar}
\affiliation{Theoretische Physik III, Universit\"at Bayreuth, 95440 Bayreuth, Germany}
\author{P. I. Tamborenea}
\affiliation{Departamento de F\'isica and IFIBA, FCEN, Universidad de Buenos Aires, Ciudad Universitaria, Pabell\'on I, C1428EHA Buenos Aires, Argentina
}
\author{V. M. Axt}
\affiliation{Theoretische Physik III, Universit\"at Bayreuth, 95440 Bayreuth, Germany}

\begin{abstract}

We explore the impact of a Rashba-type spin-orbit interaction in the conduction band on the spin dynamics of hot excitons in diluted magnetic semiconductor quantum wells.
In materials with strong spin-orbit coupling, we identify parameter regimes where spin-orbit effects greatly accelerate the spin decay and even change the dynamics qualitatively in the form of damped oscillations.
Furthermore, we show that the application of a small external magnetic field can be used to either mitigate the influence of spin-orbit coupling or entirely remove its effects for fields above a material-dependent threshold.

\end{abstract}

\maketitle

\section{Introduction}
\label{sec:Introduction}

Diluted magnetic semiconductors (DMSs) are well known for the strong exchange interaction between the localized magnetic dopants and the quasi-free carriers in the material
\cite{Bouzerar_Unified-picture_2010, Dietl_Dilute-ferromagnetic_2014, Kossut_Introduction-to_2010, Furdyna_Diluted-magnetic_1988}, an effect which famously can give rise to complete ferromagnetic ordering at sufficiently large doping concentrations in III-V compounds \cite{Ohno_Making-Nonmagnetic_1998}.
In contrast, typical II-VI DMSs are paramagnetic \cite{Furdyna_Diluted-magnetic_1988} and can be used as a spin aligner \cite{Fiederling_Injection-and_1999} or to perform spin-noise spectroscopy \cite{Cronenberger_Atomic-like_2015}.
Here, we focus on the latter material class which has the advantage that the doping atoms, typically manganese, can be incorporated isoelectronically into the host lattice so that no excess charge carriers are introduced.

Since the carrier-impurity exchange coupling is so strong, its effects typically dominate the carrier spin dynamics in DMSs for a vast range of parameters \cite{Kossut_Introduction-to_2010}.
This is why, in contrast to nonmagnetic semiconductors, other spin relaxation mechanisms such as phonon or spin-orbit effects are often not considered in theoretical models \cite{Semenov_Electron-spin_2003, Camilleri_Electron-and_2001, Morandi_Ultrafast-magnetization_2009, BenCheikh_Electron-spin_2013, Vladimirova_Dynamics-of_2008, Thurn_Quantum-kinetic_2012}.
However, it has already been shown that, for suitable materials under appropriate parameters, a nonequilibrium distribution of quasi-free electron spins can be significantly affected by spin-orbit interaction (SOI) in bulk as well as nanostructures \cite{Ungar_Ultrafast-spin_2015}.
But instead of exciting quasi-free electron spins, many experimental works choose to focus on the exciton resonance \cite{Crooker_Optical-spin_1997, Camilleri_Electron-and_2001, BenCheikh_Electron-spin_2013, Roennburg_Motional-Narrowing_2006}, which opens up the question of whether a regime of a true competition between the exchange interaction and the SOI in DMSs can be reached also in the spin dynamics of electron-hole pairs.

To answer this question, we consider a nonequilibrium distribution of so-called hot excitons under the influence of SOI.
In contrast to resonantly excited excitons, which are characterized by vanishing center-of-mass momenta due to momentum conservation, hot excitons are formed, e.g., by above band-gap excitation and subsequent relaxation onto the exciton parabola via the emission of longitudinal optical (LO) phonons \cite{Poweleit_Thermal-relaxation_1997, Umlauff_Direct-observation_1998, Chen_Exciton-spin_2003}.
This makes hot excitons ideal candidates to investigate the interplay between exchange interaction and SOI since their distribution is strongly out of equilibrium and they are created with sizable center-of-mass kinetic energies, which is favorable for SOI since it becomes stronger with larger wave numbers.

Although many articles have pointed out the importance of correlations in DMSs \cite{Ohno_Making-Nonmagnetic_1998, Thurn_Quantum-kinetic_2012, DiMarco_Electron-correlations_2013, Cygorek_Influence-of_2017, Ungar_Quantum-kinetic_2017, Ungar_Many-body_2018}, it has been recently shown that the spin dynamics of hot excitons is in fact well described by a Markovian model along the lines of Fermi's golden rule \cite{Ungar_Phonon-impact_2019}.
This is because, in DMSs, correlation effects requiring a full quantum kinetic description are most pronounced near the band edge \cite{Cygorek_Non-Markovian_2015} and the center of a hot exciton distribution is typically a few meV away from the corresponding bottom of the exciton parabola \cite{Umlauff_Direct-observation_1998}, where the density of states abruptly drops to zero.
Regarding the SOI, we focus on Rashba-type spin-orbit coupling in quasi two dimensional quantum wells in the conduction band and neglect the corresponding valence band terms.
Since the latter are inversely proportional to the splitting between heavy holes (hh) and light holes (lh) \cite{Wu_Spin-dynamics_2010}, they can be expected to be less relevant in samples with a sufficiently large hh-lh splitting such as in narrow quantum wells or in the presence of strain \cite{Winkler_Spin-Orbit_2003}.

Our findings suggest that there is indeed a regime where the SOI is comparable or even exceeds the carrier-impurity exchange interaction, provided that the doping fraction is sufficiently low and the Rashba coupling constant is large enough.
In our numerical simulations, the latter dependence is revealed by comparing the results for two different materials with drastically different band gaps and, consequently, different coupling constants, namely Zn$_{1-x}$Mn$_x$Se and Cd$_{1-x}$Mn$_x$Te.
It is also found that SOI effects are particularly pronounced for narrow quantum wells. 
Finally, we show that the influence of SOI can be drastically reduced and even be completely switched off by applying an external magnetic field, a finding which is particularly important for potential applications.

\section{Theory}
\label{sec:Theory}

In this section, the individual contributions to the Hamiltonian of a DMS quantum well with SOI in the conduction band are discussed.
We also provide the equations of motion to describe the exciton spin dynamics in the Markov limit without SOI.
To add the influence of the latter, the Rashba Hamiltonian is projected onto the exciton basis, where it is found to take the form of an effective magnetic field in terms of the exciton center-of-mass momentum that can be added to the equations of motion.

\subsection{Spin dynamics of excitons in DMSs}
\label{Spin-dynamics-of-excitons-in-DMSs}

We model the spin dynamics of hot excitons on the $1s$ exciton parabola in a II-VI DMS quantum well.
As a distribution of hot excitons typically forms on a femtosecond timescale via LO-phonon emission after above band-gap excitation \cite{Poweleit_Thermal-relaxation_1997, Umlauff_Direct-observation_1998, Chen_Exciton-spin_2003}, we do not explicitly model the exciton formation process here since it is orders of magnitude faster than the typical spin dynamics in DMSs \cite{Maialle_Exciton-spin_1993, Maialle_Exciton-spin_1994, Crooker_Optical-spin_1997, Akimoto_Coherent-spin_1997, BenCheikh_Electron-spin_2013, Kamimura_Turnover-of_2015, Kossut_Introduction-to_2010}.
Rather, we perform initial-value calculations using an exciton distribution that is close to what is reported in experiments.
Note that the distribution of excitons on the $1s$ parabola is experimentally accessible via LO-phonon assisted photoluminescence measurements \cite{Umlauff_Direct-observation_1998, Zhao_Energy-relaxation_2002, Zhao_Coherence-Length_2002}.
Restricting our model to $1s$ excitons is possible due to the energy separation between the exciton ground state and the excited states, which is $10\,$meV or larger for the systems studied here.
In combination with the finite LO-phonon energy of about $30\,$meV \cite{Umlauff_Direct-observation_1998}, this means that excited exciton states can be effectively eliminated by appropriately tuning the excess excitation energy above the band gap.

The exciton spin dynamics including the Rashba spin-orbit interaction can then be modeled by the Hamiltonian \cite{Ungar_Quantum-kinetic_2017, Ungar_Phonon-impact_2019}
\begin{align}
\label{eq:Hamiltonian}
H = H_\t{X} + H_\t{Z} + H_\t{m} + H_\t{nm} + H_\t{R}.
\end{align}
Here, $H_\t{X}$ comprises the kinetic energies of electrons and holes as well as the confinement due to the quantum well and the Coulomb interaction between the carriers.
A diagonalization of $H_\t{X}$ yields the exciton wave functions and the corresponding energies.
We also account for an external magnetic field via Zeeman terms for electrons, holes, and magnetic impurities via $H_\t{Z}$.

The interaction which typically dominates the spin dynamics in DMSs is the magnetic carrier-impurity exchange interaction \cite{Kossut_Introduction-to_2010, Thurn_Quantum-kinetic_2012, Dietl_Dilute-ferromagnetic_2014, Ungar_Quantum-kinetic_2017} denoted by $H_\t{m}$.
It comprises the spin-flip scattering of $s$-like conduction band electrons and $p$-like valence band holes with the localized electrons in the $d$ shell of an impurity ion.
Despite being typically only investigated in transport studies \cite{Kossut_On-the_1975}, recent theoretical investigations have shown that nonmagnetic scattering can also significantly affect the spin dynamics \cite{Cygorek_Influence-of_2017, Ungar_Quantum-kinetic_2017, Ungar_Trend-reversal_2018}, which is why we include nonmagnetic scattering via $H_\t{nm}$.
A more detailed description of these parts of the Hamiltonian is given in Ref.~\onlinecite{Ungar_Quantum-kinetic_2017}.

Finally, we extend the model by accounting for a Rashba-type SOI in the conduction band, the Hamiltonian of which is denoted by $H_\t{R}$ and will be discussed in more detail in the following section.
Although the scattering with longitudinal acoustic (LA) phonons leads to the eventual thermalization of hot excitons, we do not include LA phonons in our model since it has been recently shown that they have a negligible influence on the ultrafast spin dynamics of hot excitons and mainly affect the shape of the distribution \cite{Ungar_Phonon-impact_2019}.
Another mechanism which is commonly discussed in the context of exciton spin relaxation in nonmagnetic systems is the long-range exchange coupling induced by the Coulomb interaction \cite{Maialle_Exciton-spin_1993}.
For the resonant excitation of excitons with quasi-vanishing center-of-mass momenta, the typical energy associated with this interaction has been estimated previously and was found to be in the $10\,\mu$eV range, much smaller than the meV energy scale of the $sd$ exchange interaction \cite{Ungar_Trend-reversal_2018}.
Considering a distribution of hot excitons in a typical $5$-nm-wide quantum well and using the expression from Ref.~\onlinecite{Maialle_Exciton-spin_1993}, we estimate an energy of about $0.03\,$meV for the long-range exchange interaction.
This corresponds to a timescale of about $140\,$ps, which is longer than the typical timescales of the magnetic exchange interaction and the SOI for the parameters considered in section \ref{sec:Numerical-simulations} so the long-range exchange part is neglected here.

Even though our model is specifically adapted to describe the exciton spin dynamics in DMS nanostructures, it can also be straightforwardly applied to nonmagnetic systems if the magnetic exchange interaction is switched off.
In that case, also the effect of a long-range exchange interaction may become more important and should be incorporated.
Furthermore, the model could be extended to the topical material class of transition-metal dichalcogenide monolayers \cite{Brem_Exciton-Relaxation_2018}.
There, the exciton states would need to be recalculated to account for the large binding energies observed in these systems and also the valley degree of freedom can be expected to play a role \cite{Ugeda_Giant-bandgap_2014, Chernikov_Exciton-Binding_2014}.

We assume a distribution of hot excitons that is spin polarized such that electrons are in the state with $s_z = \frac{1}{2}$ and heavy holes have an angular momentum quantum number of $j_z = -\frac{3}{2}$, corresponding to the energetically lowest optically active exciton state in typical semiconductors \cite{Bastard_Wave-mechanics_1996, Winkler_Spin-Orbit_2003}.
Light-hole states with $j_z = \pm\frac{1}{2}$ are energetically separated from the hh states by the hh-lh splitting due to the confinement in the quantum well and strain \cite{Winkler_Spin-Orbit_2003}.
In the following, we consider the limit of a sufficiently large hh-lh splitting so that the hh spins remain pinned in the state $j_z = -\frac{3}{2}$ \cite{Uenoyama_Hole-relaxation_1990, Ferreira_Spin-flip_1991, Bastard_Spin-flip_1992, Crooker_Optical-spin_1997}.
Then, it is sufficient to describe the exciton spin dynamics with two states, i.e., one where the exciton-bound electron spin is oriented parallel with respect to the growth direction and another where it points in the opposite direction.
Assigning the symbol $\up$ to the former state ($s_z = \frac{1}{2}$), the latter is denoted by $\down$ ($s_z = -\frac{1}{2}$).

Without SOI, the equation of motion for the spin-up and spin-down exciton density as well as for the spin components in the quantum well plane due to the exciton-impurity interaction are given by \cite{Ungar_Quantum-kinetic_2017}
\begin{widetext}
\begin{subequations}
\label{eq:Markov-equations}
\begin{align}
\label{eq:Markov-equation-n}
\ddt n_{\K_1}^\ud =&\; 
	\frac{\pi I \NMn}{\hbar^2 V^2} \sum_{\K} \Big[ \delta\big(\omega_{\K} - \omega_{\K_1}\big) \big( n_{\K}^\ud - n_{\K_1}^\ud \big)
	\Big( \big( \Jsd^2 b^\parallel \pm 2\Jsd J_0^\t{e} b^0 + 2{J_0^\t{e}}^2 \big) F_{\etah 1s 1s}^{\etah \K \K_1}
	\nn
	& + \big( \Jpd^2 b^\parallel - 2\Jpd J_0^\t{h} b^0 + 2{J_0^\t{h}}^2 \big) F_{\etae 1s 1s}^{\etae \K \K_1}
	+ \big( 4J_0^\t{e}J_0^\t{h} - 2\Jpd J_0^\t{e} b^0 \pm 2\Jsd J_0^\t{h} b^0 \mp 2\Jsd\Jpd b^\parallel \big) F_{-\etah 1s 1s}^{\ph{-} \etae \K \K_1} \Big) 
	\nn
	& + \delta\big(\omega_{\K} - (\omega_{\K_1} \pm \omsf)\big) \Jsd^2 
	F_{\etah 1s 1s}^{\etah \K \K_1} \big( b^\pm n_{\K}^\du - b^\mp n_{\K_1}^\ud \big) \Big],
	\\
\label{eq:Markov-equation-s}
\ddt \mbf s_{\K_1}^\perp =&\; 
	\frac{\pi I \NMn}{\hbar^2 V^2} \sum_{\K} \Big[ 
	\delta(\omega_{\K} - \omega_{\K_1}) (\mbf s_{\K}^\perp - \mbf s_{\K_1}^\perp)
	\big( ( 2{J_0^\t{e}}^2 - \Jsd^2 b^\parallel ) F_{\etah 1s 1s}^{\etah \K \K_1} 
	+ ( \Jpd^2 b^\parallel + 2{J_0^\t{h}}^2 - \Jpd J_0^\t{h} b^0 ) F_{\etae 1s 1s}^{\etae \K \K_1}
	\nn
	& - ( 2\Jpd J_0^\t{e} b^0  + \Jpd J_0^\t{h} b^0 - 2J_0^\t{e} J_0^\t{h} ) F_{-\etah 1s 1s}^{\ph{-} \etae \K \K_1} \big)
	- \Big( \frac{b^-}{2} \delta\big(\omega_{\K} - (\omega_{\K_1} + \omsf)\big) 
	+ \frac{b^+}{2} \delta\big(\omega_{\K} - (\omega_{\K_1} - \omsf)\big)
	\nn
	& + 2 b^\parallel \delta(\omega_{\K} - \omega_{\K_1}) \Big) \Jsd^2 F_{\etah 1s 1s}^{\etah \K \K_1} \mbf s_{\K_1}^\perp \Big] + \bs\ome \times \mbf s_{\K_1}^\perp.
\end{align}
\end{subequations}
\end{widetext}
The overall prefactor in front of the sum contains the factor $I = 1.5$, which stems from the influence of the lowest confinement state in the quantum well under the assumption of infinitely high barriers, and the number of Mn ions $\NMn$ in the system with volume $V$.
The coupling constants $J$ are labeled according to the respective interaction:
$\Jsd$ ($\Jpd$) denotes the coupling for the $sd$ ($pd$) exchange interaction and $J_0^\t{e}$ ($J_0^\t{h}$) stems from the nonmagnetic scattering at impurities in the conduction (valence) band.
We assume a magnetic field oriented along the growth direction ($z$ axis), which enters via $\bs\ome = \frac{\gel \mu_B}{\hbar} \B + \frac{\Jsd \NMn b^0}{\hbar V} \mbf e_z$ and $\bs\omMn = \frac{\gMn \mu_B}{\hbar} \B$ for the carriers and the impurity ions, respectively.
Regarding the impurity magnetization $\mbf S$ we consider the regime of small exciton densities so the impurity spin density matrix can be described by its thermal equilibrium value.
Then, the influence of $\mbf S$ is contained in the constants $b^\pm = \frac{1}{2}(\langle \mbf S^2 - (S^z)^2 \rangle \pm \langle S^z \rangle)$, $b^\parallel = \frac{1}{2} \langle (S^z)^2 \rangle$, and $b^0 = \langle S^z \rangle$.
The spin-flip scattering shift appearing in Eqs.~\eqref{eq:Markov-equations} is given by $\hbar\omsf = \hbar\ome^z - \hbar\omMn^z$.
Finally, an analytic expression for the exciton form factors $F_{\eta_1 1s 1s}^{\eta_2 \K_1 \K_2}$ is provided in Appendix \ref{app:Analytical-solution-for-the-exciton-form-factors}.
Note that the $z$ component of the exciton spin can be obtained via $s_\K^z = \frac{1}{2}(n_\K^\up - n_\K^\down)$.

Instead of using the full quantum kinetic description of the exciton spin dynamics developed in Ref.~\onlinecite{Ungar_Quantum-kinetic_2017}, here we only consider its Markov limit.
As a recent theoretical study suggests \cite{Ungar_Phonon-impact_2019}, this is justified as long as hot excitons are considered since they are far away from the bottom of the exciton parabola where quantum kinetic effects are most pronounced \cite{Cygorek_Non-Markovian_2015}.
Note that all appearing wave vectors $\K$ are two-dimensional variables.
Since the SOI introduces an effective magnetic field that explicitly depends on the wave vector \cite{Wu_Spin-dynamics_2010}, performing an average over angles in $\K$ space to reduce the numerical demand would not capture any spin-orbit physics.
This means that even terms proportional to $\delta(\omega_{\K} - \omega_{\K_1})$ give a finite contribution to the dynamics as they only limit the sum over the absolute value $K$ but still allow for a scattering to an arbitrary angle.

\subsection{Rashba SOI in the exciton basis}
\label{subsec:Rashba-SOI-in-the-exciton-basis}

In an asymmetric quantum well, the Rashba SOI for a single electron can be written as \cite{Bychkov_Properties-of_1984, Glazov_Two-dimensional_2010, Wu_Spin-dynamics_2010}
\begin{align}
\label{eq:H_Rashba}
H_\t{R} = \alpha_\t{R} (k_y\sigma_x - k_x\sigma_y)
\end{align}
with a coupling constant $\alpha_\t{R}$ and Pauli matrices $\sigma_x$ and $\sigma_y$ that couple the electron spin with the components of the wave vector $\k$.
For the coupling constant, we use the expression \cite{deAndradaeSilva_Spin-orbit_1997, Gnanasekar_Effects-of_2006, Ungar_Ultrafast-spin_2015}
\begin{align}
\label{eq:alpha_R}
\alpha_\t{R} = \frac{\hbar^2}{2\me} \frac{\Delta}{E_\t{g}} \frac{2E_\t{g} + \Delta}{(E_\t{g} + \Delta)(3E_\t{g} + 2\Delta)} \frac{V_\t{qw}}{d}
\end{align}
with the electron effective mass $\me$, the spin-orbit splitting $\Delta$ in the valence band, the band gap $E_\t{g}$, and a potential drop $V_\t{qw}$ across a quantum well of width $d$.

In order to incorporate the SOI in our existing description of the exciton spin dynamics, we project Eq.~\eqref{eq:H_Rashba} onto the exciton basis characterized by the states $|\sigma x \K\rangle$ with a spin index $\sigma$, the exciton quantum number $x$, and the center-of-mass wave vector $\K$.
We restrict our considerations to the exciton ground state ($x = 1s$) for which the exciton wave function in the quantum well plane can be written as \cite{Bastard_Wave-mechanics_1996}
\begin{align}
\Psi_{1s \K}(\r,\R) = e^{-i\K\cdot\R} \Phi_{1s}(\r).
\end{align}
Using polar coordinates for the center-of-mass position $\R = \R(R, \phi)$ and the relative coordinate $\r = \r(r, \varphi)$, respectively, the wave function for the relative motion $\Phi_{1s}(\r) = \Phi_{1s}(r)$ of the exciton ground state does not depend on the angle $\varphi$ since an $s$ state is characterized by vanishing angular momentum.
To shorten the notation, we shall drop the index $1s$ in the following.

To express the Rashba Hamiltonian in Eq.~\eqref{eq:H_Rashba} in terms of the exciton ground state, it is convenient to first write it in terms of center-of-mass and relative coordinates as well.
This yields
\begin{align}
\label{eq:H_Rashba_spherical}
H_\t{R} = \alpha_\t{R} \big( s^+ (\partial_\R^- + \partial_\r^-) - s^- (\partial_\R^+ + \partial_\r^+)\big)
\end{align}
with the partial derivatives 
\begin{subequations}
\begin{align}
\label{eq:partial_R}
\partial^\pm_\R =&\; e^{\pm i\phi}\big(\frac{\partial}{\partial R} \pm i \frac{1}{R}\frac{\partial}{\partial\phi}\big),
	\\
\label{eq:partial_r}
\partial^\pm_\r =&\; e^{\pm i\varphi}\big(\frac{\partial}{\partial r} \pm i \frac{1}{r}\frac{\partial}{\partial\varphi}\big)
\end{align}
\end{subequations}
and spin raising and lowering operators $s^\pm$, respectively.
In second quantization with respect to the states $|\sigma \K\rangle$, one then has to compute the corresponding matrix elements of Eq.~\eqref{eq:H_Rashba_spherical}.
It turns out that matrix elements containing a derivative with respect to the relative motion vanish for the exciton ground state, i.e.
\begin{align}
\label{eq:matrix_element_r}
\langle \sigma_1 \K_1| \partial_\r^\pm | \sigma_2 \K_2 \rangle &= 0.
\end{align}
To see this, the two terms in Eq.~\eqref{eq:partial_r} can be considered separately.
First, the second part of Eq.~\eqref{eq:partial_r} containing the derivative with respect to $\varphi$ vanishes since the exciton wave function does not depend on $\varphi$.
Second, the matrix element contains an integral over $\varphi$, which vanishes due to the $\varphi$-dependent phase in Eq.~\eqref{eq:partial_r} no matter what the result of the derivative with respect to $r$ is.
For the remaining matrix element of the center-of-mass motion, it is advantageous to switch back to Cartesian coordinates, where
\begin{align}
\partial_\R^\pm = \frac{\partial}{\partial X} \pm i \frac{\partial}{\partial Y}.
\end{align}
The resulting matrix element can then be straightforwardly evaluated to be
\begin{align}
\label{eq:matrix_element_R}
\langle \sigma_1 \K_1| \partial_\R^\pm | \sigma_2 \K_2 \rangle = \alpha_\t{R} \big(-i K_{1,X} \pm K_{1,Y}\big) \delta_{\K_1,\K_2}
\end{align}
under the condition that $\Phi_{1s}(\r)$ is normalized with respect to the quantum well area.

Combining Eqs.~\eqref{eq:matrix_element_r} and \eqref{eq:matrix_element_R} with the spin selection rules enforced by $s^\pm$, the conduction-band SOI in the exciton basis becomes
\begin{align}
\label{eq:H_Rashba_exciton}
H_\t{R} =&\; \alpha_\t{R} \sum_{\K} \big( (-i K_X - K_Y) Y_{\up \K}^\dagger Y_{\down \K}
	\nn 
	&+ (i K_X - K_Y) Y_{\down \K}^\dagger Y_{\up \K} \big)
\end{align}
in terms of the creation (annihilation) operator $Y_{\sigma \K}^\dagger$ ($Y_{\sigma \K}$) of a $1s$ exciton with spin $\sigma$ and center-of-mass wave vector $\K$.
Using the Heisenberg equation of motion, Eq.~\eqref{eq:H_Rashba_exciton} leads to the typical precession-type dynamics in an effective spin-orbit magnetic field that depends on the wave vector \cite{Glazov_Nonexponential-spin_2005, Glazov_Two-dimensional_2010, Wu_Spin-dynamics_2010, Ungar_Ultrafast-spin_2015, Cosacchi_Nonexponential-spin_2017}.
For the exciton spin $\mbf s_\K$, this amounts to an additional contribution to the equation of motion of the form
\begin{align}
\left.\ddt\right|_\t{SOI} \mbf s_\K = \bs\Omega_\K \times \mbf s_\K
\end{align}
with an effective magnetic field in the quantum well plane given by
\begin{align}
\bs\Omega_\K = \frac{\alpha_\t{R}}{\hbar} \left(\begin{matrix}
	-K_Y \\
	K_X \\
	0
\end{matrix}\right).
\end{align}
This means that the SOI induces a $\K$-dependent precession of exciton spins which, provided it is strong enough, can be expected to lead to a faster spin decay due to dephasing.

Note that, in bulk semiconductors, one typically discusses the Dresselhaus SOI in combination with momentum scattering for the spin relaxation, which is known as the Dyakonov-Perel mechanism \cite{Dyakonov_Spin-Physics_2008}.
However, in two dimensional systems such as quantum wells, the strong confinement in the $z$ direction reduces the characteristic cubic dependence of the Dresselhaus spin-orbit field on $\k$ to a linear dependence, just like in Eq.~\eqref{eq:H_Rashba} but with a different coupling constant and proportional to $k_x \sigma_x - k_y \sigma_y$.
In contrast to the Rashba interaction, where $\alpha_\t{R}$ is tunable via an electric field, the Dresselhaus coupling constant is fixed and depends only on the specific semiconductor \cite{Winkler_Spin-Orbit_2003}.
As discussed in the previous section, momentum scattering is present in our model due to the scattering of excitons at the impurities in the DMS.
Thus, provided that the coupling constants are comparable, a similar spin decay is expected due to the Rashba and the Dresselhaus SOI independently.
Although including a Dresselhaus term in our model is straightforward, studying this mechanism is beyond the scope of this paper.

\section{Numerical simulations}
\label{sec:Numerical-simulations}

To analyze how SOI in the conduction band can impact the exciton spin dynamics of hot excitons in DMS quantum wells, we perform numerical simulations for two different materials, namely Zn$_{1-x}$Mn$_x$Se and Cd$_{1-x}$Mn$_x$Te, for varying doping fractions $x$ and different quantum well widths.
Furthermore, we explore the impact of an external magnetic field.
We assume an initial distribution of hot excitons on the $1s$ parabola that is spin-polarized in the $\up$ state.
The distribution is modeled as a Gaussian with a standard deviation of $1\,$meV centered at $10\,$meV above the bottom of the $1s$ exciton parabola, which is similar to what has been observed in experiments for ZnSe \cite{Umlauff_Direct-observation_1998}.
The spin moments of the impurity ions are calculated using a fixed temperature of $2\,$K.

The microscopic parameters used for the simulations are the same as in Ref.~\onlinecite{Ungar_Quantum-kinetic_2017}.
To calculate the SOI prefactor given by Eq.~\eqref{eq:alpha_R}, we use a spin-orbit splitting $\Delta = 403\,$meV and a band gap $E_\t{g} = 2.820\,$eV for Zn$_{1-x}$Mn$_x$Se,
while $\Delta = 949\,$meV and $E_\t{g} = 1.606\,$eV are used for Cd$_{1-x}$Mn$_x$Te \cite{Winkler_Spin-Orbit_2003}.
Apart from the coupling constants $J_0^\t{e}$ and $J_0^\t{h}$, which are obtained by considering the change of the band gap when going over from a completely undoped semiconductor to a compound where all group II atoms have been replaced by Mn \cite{Cygorek_Influence-of_2017}, all other microscopic parameters are taken directly from experiments.
Since we study DMSs with a relatively small concentration of doping atoms, the change of the band gap with increasing doping fraction is disregarded here.
In all simulations, a potential drop $V_\t{qw} = 100\,$meV across the quantum well is assumed.
Considering, e.g., a $10$-nm-wide DMS quantum well, we obtain $\alpha_\t{R} \approx 0.07\,$meV$\,$nm for Zn$_{1-x}$Mn$_x$Se and $\alpha_\t{R} \approx 0.36\,$meV$\,$nm for Cd$_{1-x}$Mn$_x$Te, respectively.
These values are realistic compared to what has been reported in the literature \cite{Schliemann_Nonballistic-Spin_2003, Bindel_Probing-variations_2016}.
Note that the Rashba coefficient is recalculated for simulations where the quantum well width is varied using the constant value for the potential drop across the well given above.

\begin{figure}
	\centering
	\includegraphics{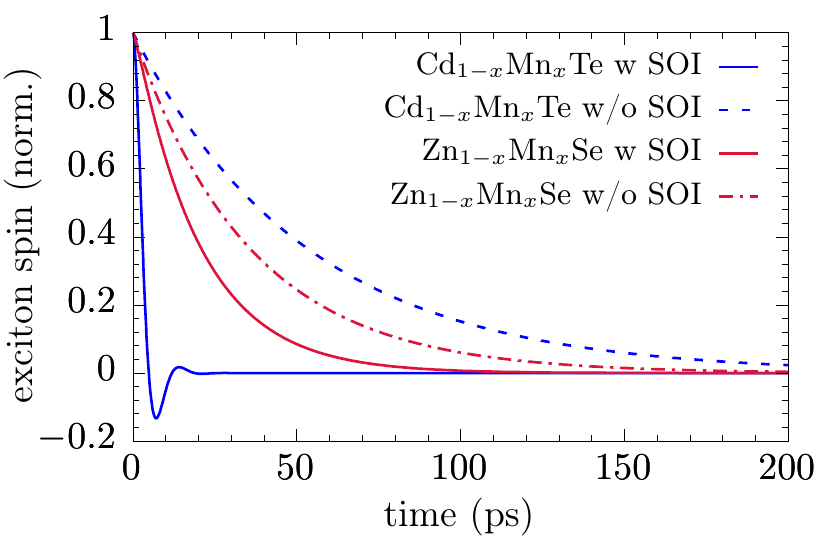}
	\caption{Spin dynamics in $5$-nm-wide DMS quantum wells with a doping fraction $x = 0.2\%$.
	Results are shown for two different compounds with and without spin-orbit interaction (SOI)
	taken into account.}
	\label{fig:dynamics}
\end{figure}

Since the SOI has to compete with the rather strong carrier-impurity exchange interaction in DMSs, effects of the former can be expected to become particularly relevant for small impurity content where the exchange interaction is less significant \cite{Ungar_Ultrafast-spin_2015}.
Indeed, Fig.~\ref{fig:dynamics} shows that the SOI can have a substantial impact on the spin dynamics of hot excitons in $5$-nm-wide DMS quantum wells with $0.2\%$ Mn doping.
In the case of Cd$_{1-x}$Mn$_x$Te, the calculation including the SOI reveals a substantially faster spin decay accompanied by an oscillatory behavior on short time scales compared to the calculation without SOI, which shows a standard exponential decay.
For Zn$_{1-x}$Mn$_x$Se, the SOI also accelerates the decay but does not lead to any oscillations.
Compared to the previous compound, its impact is much less pronounced.

Since a Markovian description of the carrier-impurity exchange interaction results in a rate-type dynamics \cite{Rossi_Theory-of_2002}, a monoexponential spin decay is expected on that level of theory \cite{Wu_Spin-dynamics_2010}.
Figure~\ref{fig:dynamics} reveals that this behavior can change dramatically when the SOI becomes significant, leading to a spin decay which is no longer exponential and may even show oscillations.
In fact, a nonexponential decay is a characterisic feature of the presence of SOI\cite{Glazov_Nonexponential-spin_2005, Ungar_Ultrafast-spin_2015, Cosacchi_Nonexponential-spin_2017}.
As noted before, the impact of the SOI on the spin dynamics is much stronger for Cd$_{1-x}$Mn$_x$Te than for Zn$_{1-x}$Mn$_x$Se for a given doping fraction and nanostructure.
The reason for this is the larger Rashba coupling constant in the former compound due to its significantly smaller band gap.
A typical time scale for the SOI can be roughly obtained by $\tau_\t{SOI} \approx \frac{h}{\alpha_\t{R} \bar{K}}$, where $\bar{K}$ is the average exciton wave number.
This definition is such that $\tau_\t{SOI}$ corresponds to the precession time in the effective magnetic field due to the SOI at the fixed wave number $\bar{K}$.
Using the parameters of Fig.~\ref{fig:dynamics} with a value of $\bar{K}$ corresponding to the center of the exciton distribution, we find $\tau_\t{SOI} \approx 12.4\,$ps and $\tau_\t{SOI} \approx 56.4\,$ps for Cd$_{1-x}$Mn$_x$Te and Zn$_{1-x}$Mn$_x$Se, respectively, which fits well to the behavior of the curves with SOI taken into account and also confirms that the SOI in Cd$_{1-x}$Mn$_x$Te is about five times stronger than in Zn$_{1-x}$Mn$_x$Se.

The reason for the accelerated decay of the $z$ component of the exciton spin in the presence of SOI is the $\K$-dependent magnetic field, which causes a precession whose frequency depends not only on the absolute value of the wave vector but also on its angle.
This causes individual spins in an ensemble to dephase and, when looking at the average spin in the system, leads to a decay \cite{Wu_Spin-dynamics_2010}.
As shown in Fig.~\ref{fig:dynamics}, SOI effects are particularly pronounced for hot excitons since their distributions spans a wide range of $\K$ vectors with sizable absolute values.
This is in stark contrast to optically excited excitons which are generated with quasi vanishing wave vectors close to the exciton resonance, where SOI effects are thus expected to be much less significant.

\begin{figure}
	\centering
	\includegraphics{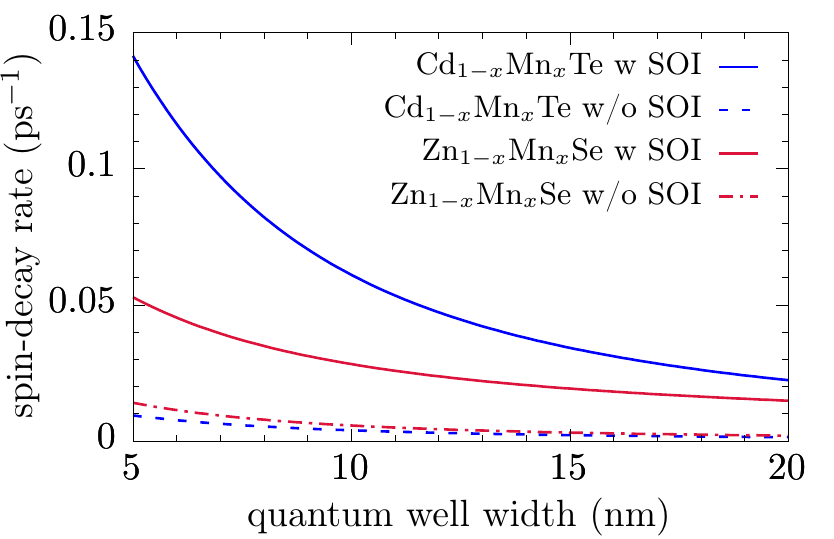}
	\caption{Spin-decay rate in DMS quantum wells with a doping fraction $x = 0.1\%$ as a function of the 
	well width.
	Results are shown for two different compounds with and without spin-orbit interaction (SOI)
	taken into account.}
	\label{fig:dvar_0.1}
\end{figure}

Without an external magnetic field and SOI, a simple $K$-dependent expression for the spin-decay rate due to the $sd$ exchange interaction can be derived \cite{Ungar_Quantum-kinetic_2017}.
For the parameters of Fig.~\ref{fig:dynamics}, one then obtains a corresponding spin relaxation time $\tau_{sd} \approx 41.0\,$ps and $\tau_{sd} \approx 26.0\,$ps for Cd$_{1-x}$Mn$_x$Te and Zn$_{1-x}$Mn$_x$Se, respectively, when the expression is evaluated with the value $\bar{K}$ used previously for the SOI.
These numbers provide the correct time scales seen in the spin decay in Fig.~\ref{fig:dynamics} without SOI, although one has to be aware that, since the distribution of hot excitons extends over a wide $K$ range, the correct spin relaxation time would have to be obtained by averaging the $K$ dependence of the rate weighted according to the exciton distribution.
In any case, comparing the time scales of the SOI with those of the exchange interaction nicely shows that we are in a regime where spin-orbit coupling competes with or even dominates the magnetic scattering.

To find out if such a regime can also be reached in larger nanostructures, we plot the spin-decay rate as a function of the quantum well width in Fig.~\ref{fig:dvar_0.1} for DMSs with a doping fraction $x = 0.1\%$.
The spin-decay rate is obtained numerically as the inverse time where the spin has decayed to $\frac{1}{e}$ with respect to its initial value.
If oscillations appear in the spin dynamics, the envelope of a decaying cosine is used for the extraction of the spin relaxation time so as not to capture the oscillations themselves in the spin-decay rate.
The same procedure is used whenever a spin-decay rate is determined in the following.

It turns out that the SOI accelerates the spin decay even for relatively large quantum wells with a width of $20\,$nm.
As found before, this increase is stronger for Cd$_{1-x}$Mn$_x$Te compared with Zn$_{1-x}$Mn$_x$Se and can be larger than an order of magnitude for very small quantum wells in the former compound.
We find an inverse dependence of the rate on the quantum well width for the results with and without SOI, which is the common tendency in DMSs \cite{Bastard_Spin-flip_1992, Maialle_Exciton-spin_1993, Akimoto_Carrier-spin_1997}.
The inverse dependence on the width can be directly inferred from the Rashba prefactor given by Eq.~\eqref{eq:alpha_R} as well as the spin-decay rate in the absence of SOI \cite{Ungar_Quantum-kinetic_2017}.

\begin{figure}
	\centering
	\includegraphics{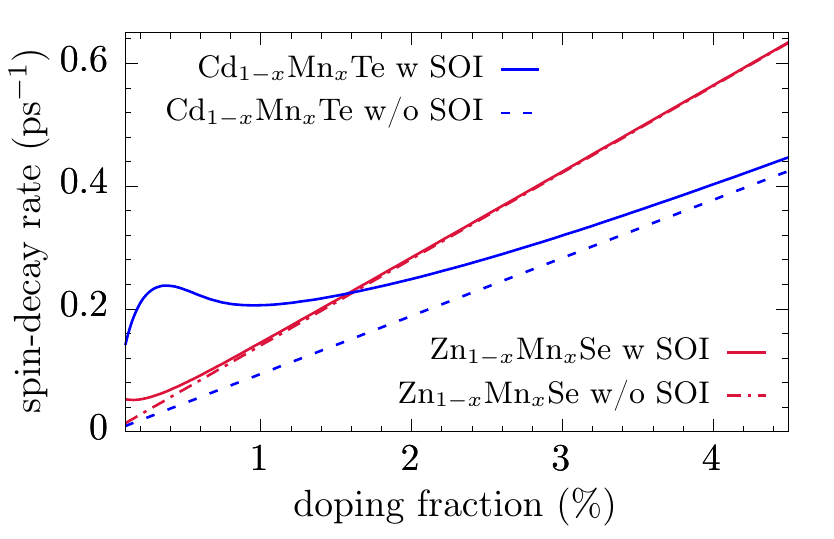}
	\caption{Spin-decay rate in $5\,$-nm-wide DMS quantum wells as a function of the doping fraction $x$.
	Results are shown for two different compounds with and without spin-orbit interaction (SOI)
	taken into account.}
	\label{fig:xMnvar}
\end{figure}

However, in contrast to the SOI, the time scale of the exchange interaction is strongly controlled by the amount of impurities in the sample \cite{Kossut_Introduction-to_2010, Ungar_Trend-reversal_2018}, suggesting that a dominance of SOI effects should disappear for larger doping fractions \cite{Ungar_Ultrafast-spin_2015}.
Indeed, as shown in Fig.~\ref{fig:xMnvar}, the impact of the SOI on the spin-decay rate decreases with increasing impurity content.
For Zn$_{1-x}$Mn$_x$Se, where Rashba spin-orbit coupling is rather weak, we find that the exchange interaction clearly dominates the exciton spin decay for doping fractions above $0.5\%$.
A similar behavior was found for the spin dynamics of quasi-free electrons in the same material \cite{Ungar_Ultrafast-spin_2015}.
However, the spin dynamics in Cd$_{1-x}$Mn$_x$Te remains visibly affected by the SOI even at doping fractions of a few percent, suggesting that SOI effects may contribute to the experimentally determined significant spread of spin relaxation rates for different samples collected in Ref.~\onlinecite{BenCheikh_Electron-spin_2013}.

While the results for Zn$_{1-x}$Mn$_x$Se with SOI go smoothly over to those without SOI, there is a visible local maximum in the spin-decay rate for small doping fractions below $1\%$ in Cd$_{1-x}$Mn$_x$Te.
There, the rate obtained when the SOI is included first rises with increasing Mn content, then decreases and finally increases once more.
The second increase starting at about $1\%$ doping stems from the exchange interaction, which is directly proportional to the number of Mn ions in the sample [cf. Eqs.~\eqref{eq:Markov-equations}] as can be seen from the behavior of the curve without SOI.
In contrast, the initial rise, maximum and following decrease of the rate is indicative of a qualitative change in the dynamics.
For very small doping fractions where the SOI completely dominates, the $z$ component of the spin exhibits a decaying oscillatory behavior (cf. Fig.~\ref{fig:dynamics}).
This is eliminated with increasing strength of the exchange interaction as the latter always leads to an exponential decay without any oscillations.
In that sense, the local maximum observed in Fig.~\ref{fig:xMnvar} is related to a change from the oscillatory to the nonoscillatory regime that takes place for small doping fractions.

\begin{figure}
	\centering
	\includegraphics{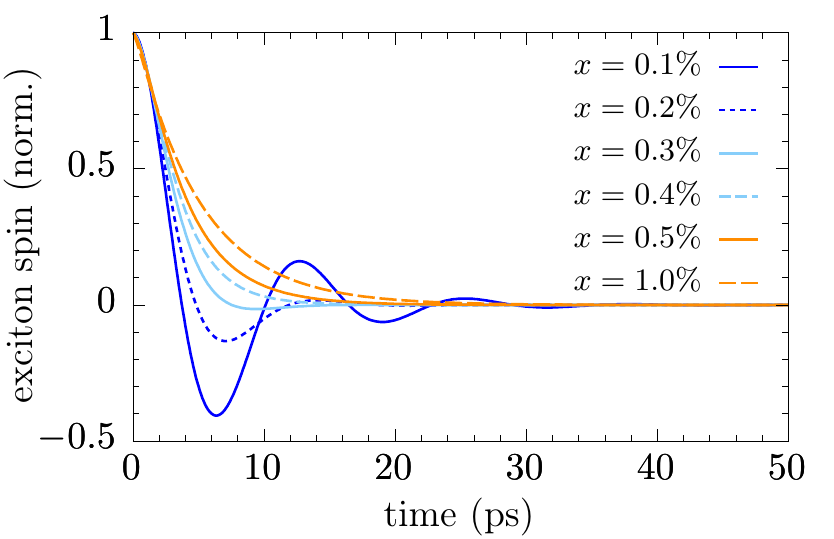}
	\caption{Spin dynamics in a $5$-nm-wide Cd$_{1-x}$Mn$_x$Te quantum well for varying doping fractions $x$ under the influence of SOI.}
	\label{fig:CdTe_dynamics}
\end{figure}

To corroborate this interpretation, we show the exciton spin dynamics in a $5$-nm-wide Cd$_{1-x}$Mn$_x$Te quantum well under the influence of SOI for small doping fractions in Fig.~\ref{fig:CdTe_dynamics}.
While below doping fractions of about $0.3\%$ pronounced oscillations can be observed, they disappear entirely above that threshold and the dynamics becomes exponential.
As stated before, the appearance of oscillations in the spin dynamics is a tell-tale sign of spin-orbit effects, so that we find indeed that the spin dynamics is dominated by SOI at very small doping fractions and a regime change occurs when the doping increases.
The value of the threshold deduced from Fig.~\ref{fig:CdTe_dynamics} also nicely fits to the position of the local maximum in Fig.~\ref{fig:xMnvar}.
In contrast, the spin dynamics in Zn$_{1-x}$Mn$_x$Se never reaches the point where it is dominated by SOI so that oscillations appear, not even at the smallest doping fractions considered here (cf. Fig.~\ref{fig:dynamics}).

\begin{figure}
	\centering
	\includegraphics{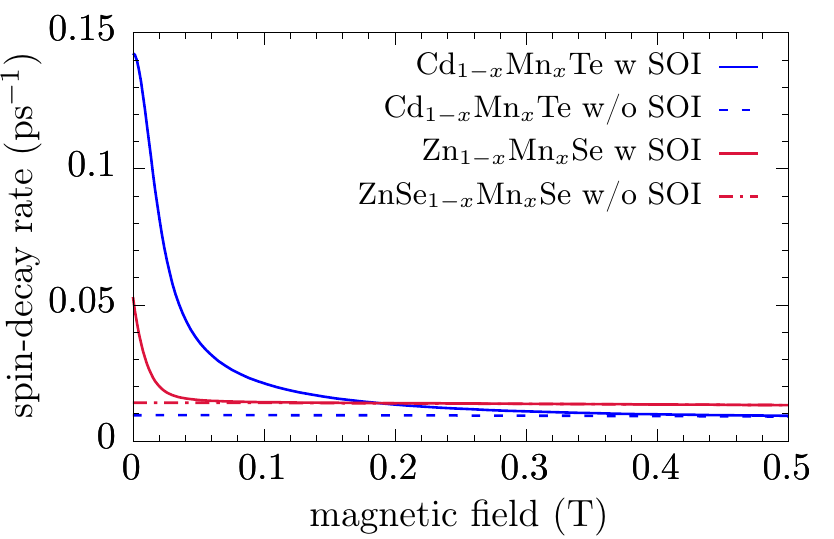}
	\caption{Spin-decay rate in $5\,$-nm-wide DMS quantum wells with a doping fraction $x = 0.1\%$ as a
	function of the applied external magnetic field.
	Results are shown for two different compounds with and without spin-orbit interaction (SOI)
	taken into account.}
	\label{fig:Bvar}
\end{figure}

Since our results reveal a strong impact of SOI on the spin dynamics for a range of parameters, the question arises if its effects can be mitigated or even entirely removed.
It turns out that this can be achieved by applying an external magnetic field, as shown in Fig.~\ref{fig:Bvar}.
Choosing parameters where spin-orbit coupling was found to be most pronounced, i.e., small well widths and small doping fractions, the increased spin-decay rate due to the SOI quickly drops to the rate without SOI when an external magnetic field is applied.
When the field reaches approximately $50\,$mT in the case of Zn$_{1-x}$Mn$_x$Se and $300\,$mT in the case of Cd$_{1-x}$Mn$_x$Te, any increase due to the SOI has almost entirely disappeared.
Thus, unwanted SOI effects can be easily suppressed by applying a relatively moderate magnetic field.

The reason why small external magnetic fields are already sufficient to inhibit the SOI is the giant Zeeman splitting in DMSs \cite{Furdyna_Diluted-magnetic_1988, Kossut_Introduction-to_2010, Dietl_Dilute-ferromagnetic_2014}, which enhances the local magnetic field experienced by the carriers.
Since the giant Zeeman splitting is so large in these structures, it quickly becomes the dominant contribution with increasing magnitude of the applied field so that, in comparison, the effective magnetic field due to the SOI becomes less and less significant.
This is also the reason why the impact of SOI on the spin dynamics of hot excitons has, to the best of our knowledge, not been observed in experiments so far as most of the experimental data available for the spin dynamics of hot excitons in DMS nanostructures has been obtained in finite magnetic fields \cite{Chen_Exciton-spin_2003, Buyanova_Resonant-suppression_2003, Buyanova_Effect-of_2005}.
On the other hand, studies performed without magnetic field have been focused so far typically on the exciton thermalization dynamics and did not probe the spin \cite{Poweleit_Thermal-relaxation_1997, Umlauff_Direct-observation_1998, Zhao_Energy-relaxation_2002}.

\section{Conclusion}
\label{sec:Conclusion}

We have derived and implemented Rashba spin-orbit coupling for excitons in DMS quantum wells to
investigate its impact on the spin dynamics of hot excitons in these materials.
The hot excitons are modeled using an initial distribution of excitons on the $1s$ parabola that is chosen according to measurements performed in ZnSe quantum wells after optical excitation above the band gap and subsequent LO-phonon relaxation \cite{Umlauff_Direct-observation_1998}.
Numerical simulations have been performed for two compounds, namely Zn$_{1-x}$Mn$_x$Se and Cd$_{1-x}$Mn$_x$Te, for which spin-decay rates were extracted and compared.

It is found that the SOI is particularly pronounced for small doping fractions and narrow quantum wells, where it leads to a sizable increase of the spin-decay rate compared with simulations not accounting for spin-orbit coupling.
The faster spin decay observed in the simulations is a consequence of dephasing in the wave-vector dependent effective magnetic field provided by the SOI.
In Cd$_{1-x}$Mn$_x$Te quantum wells, where the SOI is especially strong due to the relatively small band gap and high spin-orbit splitting, SOI effects even cause visible oscillations in the spin dynamics for small doping fractions which disappear when the impurity content increases.
The appearance of a local maximum in the spin-decay rate as a function of the doping fraction is indicative of this qualitative change from a decaying oscillatory to an exponentially decaying dynamics.
In the bulk limit as well as for samples with higher impurity content, our results suggest that SOI effects are either completely suppressed or at least strongly reduced, which is in line with previous works \cite{Ungar_Ultrafast-spin_2015}.

Finally, we have shown that the influence of SOI on the spin dynamics in DMSs can be overcome by applying a moderate external magnetic field.
Due to the strength of the giant Zeeman effect in these materials, the exchange interaction quickly becomes the dominant spin-decay mechanism as soon as the magnetic field passed a threshold determined by the specific material.
Experimentally, applying a small magnetic field may thus be used to mitigate unwanted spin-orbit contributions.

\section{Acknowledgements}
\label{sec:Acknowledgements}

We gratefully acknowledge the financial support of the Deutsche Forschungsgemeinschaft (DFG) through Grant No. AX17/10-1.

\section*{Appendix: Analytical solution for the exciton form factors}
\label{app:Analytical-solution-for-the-exciton-form-factors}

The wave-vector dependent exciton form factors can be calculated via the relation\cite{Ungar_Quantum-kinetic_2017}
\begin{align}
\label{eq:form-factor-F}
F_{\eta_1 1s 1s}^{\eta_2 \K_1 \K_2} = f_{\eta_1 1s 1s}^{\ph{\eta_1} \K_1 \K_2} \big(f_{\eta_2 1s 1s}^{\ph{\eta_2} \K_1 \K_2}\big)^*
\end{align}
with
\begin{align}
\label{eq:form-factor-f}
f_{\eta 1s 1s}^{\ph{\eta} \K_1 \K_2} = 2\pi \int_0^\infty dr \, r \phi_{1s}^2(r) J_0\big(\eta r |\K_1-\K_2|\big),
\end{align}
where $\eta$ is either the ratio between the hh or the electron mass and the exciton mass.
A common ansatz for a trial exciton wave function is given by \cite{Bastard_Spin-flip_1992, Bastard_Wave-mechanics_1996}
\begin{align}
\label{eq:phi_1s}
\phi_{1s}(r) = \frac{\beta_{1s}}{\sqrt{2\pi}} e^{-\frac{1}{2}\beta_{1s}r}
\end{align}
with a free parameter $\beta_{1s}$ that is typically determined using a variational approach.
Here, we use $\beta_{1s}$ as a fitting parameter which is chosen such that an optimal fit to the numerical solution of the exciton problem is obtained, whereas the exciton problem itself is solved in real space using a finite-difference method.
Substituting Eq.~\eqref{eq:phi_1s} into Eq.~\eqref{eq:form-factor-f} allows one to evaluate the appearing integral analytically, which yields
\begin{align}
\label{eq:form-factor-f-integrated}
f_{\eta 1s 1s}^{\ph{\eta} \K_1 \K_2} = \frac{1}{\Big( 1 + \big(\frac{\eta |\K_1 - \K_2|}{\beta_{1s}}\big)^2\Big)^\frac{3}{2}}.
\end{align}
Finally, inserting this into Eq.~\eqref{eq:form-factor-F}, we obtain a closed expression for the exciton form factor that only depends on $\beta_{1s}$.

\bibliography{references}
\end{document}